\documentclass{evn2004}
\usepackage{txfonts}
\begin{document}
\title{The two sided parsec scale structure of the Low Luminosity
  Active Galactic Nucleus in NGC~4278}
\author{G. Giovannini\inst{1,2}
        \and
M. Giroletti\inst{1,2}
        \and
        G. B. Taylor\inst{3}}

\institute{Istituto di Radioastronomia, CNR/INAF, via Gobetti
  101, 40129, Bologna, Italy
  \and
Dipartimento di Astronomia, Universit\`a di Bologna,
  via Ranzani 1, 40127 Bologna, Italy
\and
National Radio Astronomy Observatory, P.O. Box O,
  Socorro, NM 87801, USA}

   \abstract{
We present new Very Long Baseline Interferometry observations of the
LINER galaxy NGC 4278 with
a linear resolution of $\la 0.1$~pc. Our radio data
reveal a two sided structure, with symmetric $S$-shaped jets emerging
from a flat spectrum core.
By comparing the positions of the components in
two epochs, we measure motions corresponding to apparent velocities
$\la 0.2\,c$, and to
ages in the range $8.3 - 65.8$~years. From our measurements, we derive
that NGC~4278 has mildly relativistic jets ($\beta \sim 0.75$),
closely aligned to the line-of-sight ($2^{\circ} \le \theta \le
4^{\circ}$). We also present a flux density history for the source with data
between 1972 and 2003.
All these arguments indicates that the low power radio emission from
NGC~4278 is emitted via the synchrotron process by relativistic
particles accelerated by a supermassive black hole.
   }

   \maketitle
%
%________________________________________________________________

\section{Introduction}

Although objects hosting an Active Galactic Nucleus represent only a
small fraction of the total number of extragalactic sources, there is
growing evidence that some kind of nuclear activity at lower level
might be a much more common feature among galaxies. Objects presenting
spectral signature of such activity include low-ionization nuclear
emission-line region, low luminosity Seyfert
galaxies, and ``transition nuclei'', i.e. nuclei with spectra
intermediate between LINERs and HII regions. These objects are grouped
under the name of Low Luminosity Active Galactic Nuclei (LLAGN, Ho et al. 
1997).
 
NGC 4278 is a nearby LLAGN. It has been investigated in detail
at most wavelengths. In the optical, HST observations reveal a central
point source and a large distribution of dust located north-northwest
of the core (Carollo et al. 1997). Ionized nuclear gas typical of LINER is
found in this galaxy by \cite{gou94}, possibly associated to a more
external ring of neutral hydrogen in PA 135$^{\circ}$ (Raimond et al. 1981).

Radio continuum observations on kpc scale have been carried
out at frequencies between 5 and 43 GHz, revealing a compact
source (Di Matteo et al. 2001; Nagar et al. 2000, 2001). 
Compactness and flat radio spectra suggest that the
radio emission is non thermal, and in this respect the emission from
NGC4278 seems to be very similar to that of powerful radio loud AGNs,
such as QSO and BL Lacs; however, the total radio luminosity of the
source is only $P_\mathrm{1.4 GHz} = 10^{21.6}$ W Hz$^{-1}$, i.e. at
least two orders of magnitude less than those powerful objects.

On parsec scale, early VLBI experiments at 18 cm and 6 cm have
revealed a core dominated structure, with an elongated feature
extending to the north-west and possibly to the south on scales of
some 10 mas (Jones et al. 1981, 1982, 1984; Schilizzi et al. 1983).
More recent observations
with the VLBA at 6 cm reveal an extended core and an
elongated region to the southeast on scales of a few milliarcsecond
(Falcke et al. 2000). \cite{gio01} 
thanks to more short spacings provided by a VLA antenna
in addition to the full VLBA, have also detected emission on the
opposite side of the core.
\cite{bon04} also detect a trace of two-sided
emission, although heavily resolved.  Finally, VLBA phase-referenced
observations have succeeded in detecting the source on sub-pc scale
even at 43 GHz, but showing only a core and a hint of low level
emission to the north (Ly et al. 2004).
 
In the present paper, we consider new VLBA observations at 5 and 8.4 GHz, 
taken on 2000 August 27
and compare the new 5
GHz image to Giovannini et al. (2001) data, 
discussing the morphology and the motion
of components. For more details we refer to Giroletti et al. (2004). 
 
NGC~4278 has a direct distance measurement of 14.9 Mpc
(Jensen et al. 2003). At this distance, 1 mas corresponds to a linear 
scale of 0.071 pc.  We
define the spectral index $\alpha$ following the convention that
$S(\nu) \propto \nu^{-\alpha}$.

\section{Observations and Data Reduction}

We observed NGC4278 at 5 and 8.4 GHz, with an 11 element 
VLBI array composed by the NRAO
Very Long Baseline Array (VLBA) and a single 25~m VLA antenna for 10
hours.
The observing run was performed on 2000 August 27, switching
between 5 GHz and 8.4 GHz. 

The correlation was carried out at the AOC in Socorro. The
distribution tapes were read into the NRAO Astronomical Image
Processing System (AIPS) for the initial calibration and the two
frequencies were separated. After that, we followed the same scheme
for the data reduction of both 5 GHz and 8.4 GHz data-sets. As a
first step, we corrected our data entering the accurate position
information obtained by \cite{ly04} (RA 12$^h$ 20$^m$ 06$^s$.825429, Dec
$29^{\circ}$ $16\arcmin$ $50\arcsec.71418$). We then performed the
usual calibration stages (removal of instrumental single band delay,
of phase and delay $R-L$ offsets, and bandpass calibration) using
scans on 3C279.
Thanks to the good data calibration and position information, we could
obtain final images with only a few iterations of phase
self-calibration.  One cycle of amplitude self-calibration with a long
solution interval (30 minutes) has also been performed before
obtaining the final $(u,v)-$data.

We also re-analyzed VLBA+Y1 5 GHz data obtained in 1995 (22 July),
taking advantage of the new position (Ly et al. 2004).  

\section{Results}
 
The final images reveal a source dominated by a central compact
component, with emission coming from either side (Figs. 1 and 2). 

   \begin{figure}
   \centering
   \vspace{220pt}
   \includegraphics{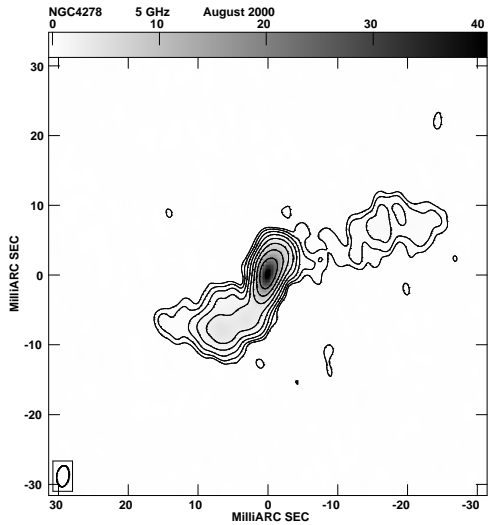}
      \caption{VLBA+Y1 image of NGC4278 at 5 GHz. Contours are at 
(1, 2, 4, ..., 128) times the lowest contour which is 0.15 mJy/beam. 
The HPBW is 3$\times$1.7 mas in PA -7$^\circ$}
   \end{figure}

   \begin{figure}
   \centering
   \vspace{220pt}
   \includegraphics{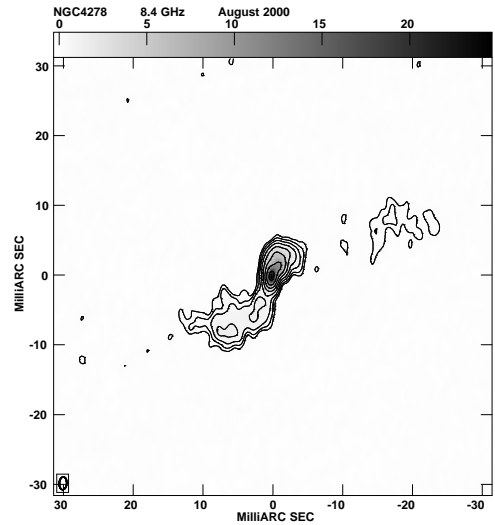}
      \caption{VLBA+Y1 image of NGC4278 at 8.4 GHz. Contours are at 
(1, 2, 4, ..., 128) times the lowest contour which is 0.15 mJy/beam. 
The HPBW is 1.8$\times$1.0 mas in PA -4$^\circ$}
   \end{figure}

   \begin{figure}
   \centering
   \vspace{160pt}
   \includegraphics{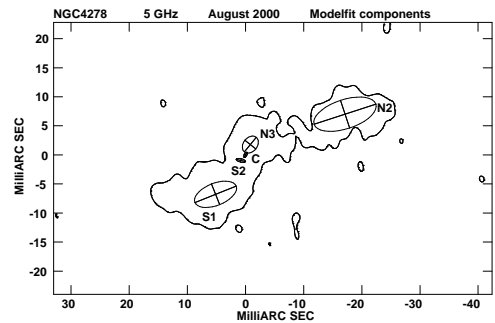}
      \caption{Model components for epoch 2000.65, overlayed to the lowest
contour from the 5 GHz image.}
   \end{figure}

To the southeast, a
jet-like feature extends for $\sim 6.5$ mas in PA 155$^{\circ}$
(measured north to east), then progressively bends into PA
100$^{\circ}$. In total, the jet is almost 20 mas long, which
corresponds to $\sim 1.4$ pc. On the opposite side, the main component
is slightly elongated to the north in the 5 GHz map (Fig. 1), and the 8.4 GHz
data clearly show a secondary component in PA $-40^{\circ}$ (Fig. 2). Then,
this jet-like feature bends to the west turning into a diffuse,
uncollimated, low brightness emitting region.
In total, the source extends over $\sim 45$ milliarcsecond, i.e. about
3 parsec.
The total flux density measured in our images is 120 mJy at 5 GHz and
95 mJy at 8.4 GHz. If we compare these values with those obtained with
the VLA in nearby epochs (162 mJy and 114 mJy, respectively), 
we find about a 20-25\% offset, which can be ascribed to the VLBA
resolving out some extended emission, probably in the western
region. The monochromatic luminosity at 1.4 GHz is $3.18 \times
10^{21}$ W Hz$^{-1}$, and $2.52 \times 10^{21}$ W Hz$^{-1}$ at 8.4
GHz.

The visibility data are well fitted by a five component model at both
frequencies. The position and dimension of the components are
illustrated in Fig. 3; our choice for labeling the
components is based on their most likely epoch of ejection, as
discussed in \S~\ref{motion}. 
Besides
being the most compact feature, component $C$ presents also the
flattest spectral index ($\alpha = 0.2$) and its identification with
the core is straightforward.
 
The same five component model that fits the 2000 epoch data has been
applied to 1995 data set as well, allowing for the components to
change in flux and position. 
 
\subsection{Component motion}
\label{motion}
 
If we compare data taken at the same frequency in different epochs, we
can get information on the evolution of the source.
Taking
the position of the core (component $C$) as a reference, and assuming
it is fixed, we have compared the position of the other components. We
report the results in Table 1: column (1) labels the
components, column (2) report the apparent velocity in 
units of $c$, and the corresponding age in years is given in
column (3). The radial distance of each component has increased over
the five years lag between the observations. The motion is larger in
the northwestern side; in particular, the largest displacement is
found for component $N2$. 

\begin{table}
\caption{Component Motion at 5 GHz}             % title of Table
\label{table:1}      % is used to refer this table in the text
\centering                          % used for centering table
\begin{tabular}{c c c}        % centered columns (3 columns)
\hline\hline                 % inserts double horizontal lines
Component & $\beta_{app}$ & age (yrs) \\
\hline                        % inserts single horizontal line
C  & reference & \\
S2 &$0.020\pm0.006$&$29.1\pm 9.3$ \\
S1 & $0.030 \pm 0.006$ & $65.8\pm 12.4$ \\
N3 & $0.055 \pm 0.004$ & $8.3\pm 0.5$ \\
N2 & $0.171 \pm 0.030$ & $25.0\pm 4.8$ \\

\hline                                   %inserts single line
\end{tabular}
\end{table}
 
Assuming that the apparent velocity is constant for each component, we
derive ages as reported in column~(3) with respect to the 2000.652 epoch. 
$S2$ and $N2$ have ages that
are consistent, and they must have been ejected together about 25
years before epoch 2000.652. $S1$ is the oldest component, and 
its counterpart in the
main jet is not detected, probably being too distant and
extended. Finally, $N3$ is the youngest component, ejected only three
years before our first epoch of observations; it is likely that a
corresponding component $S3$ has emerged in the counterjet but that it
is still confused with the core. Note that the core is the only
component whose flux density is larger in 2000.652 than in 1995.551.
 
\section{Discussion}
 
\subsection{Jet orientation and velocity}
\label{jets}
 
Our images detect low level emission to the northwest with
unprecedented resolution and sensitivity, both in the inner part of
the jet, revealing the compact component N3, and at a larger distance,
detecting $\sim 10$ mJy of flux in the region N2. Thus, we classify
NGC 4278 as a two-sided source, similarly to a few other LLAGN
previously studied, e.g. NGC4552 (Nagar et al. 2002), NGC 6500
(Falcke et al. 2000), and NGC 3894 (Taylor et al. 1998).

Although the southern jet looks
more collimated, the total flux density is larger in the northern
components than in the southern ones. Moreover the high resolution 8.4 GHz
image (Fig. 2) clearly shows that the inner jet is brighter
in its northern part than to the south, and the apparent motion of the
northern components are also larger than those of the southern
ones. Therefore we assume that the main and approaching jet is the 
northern one.
 
To estimate the orientation $\theta$ and intrinsic velocity $\beta$ of
the jet, we will consider a simple beaming model, which assumes that
components are ejected in pairs from the core at the same time, with
the same intrinsic velocity and brightness; we apply this model to the
components pair $N2$/$S2$, which have been ejected simultaneously
according to our motion measurments (Table 1). In this model, the ratio
between the arm length $r$ and the proper motion $\mu$ of the two
components are related by
 
%$R = \mu_{N2}/\mu_{S2} = r_{N2}/r_{S2} = (1+\beta\cos\theta)/(1-\beta\cos\theta)$
 
$$R = \frac{\mu_{N2}}{\mu_{S2}} = \frac{r_{N2}}{r_{S2}} = \frac{1+\beta\cos\theta}{1-\beta\cos\theta}$$
 
From our modelfits we derive that $r_{N2}/r_{S2} = 7.2 \pm 0.2$ and
$\mu_{N2}/\mu_{S2} = 8 \pm 3$, so we estimate that $4 < R < 10$. The
arm length ratio corresponds to selecting the hatched area between the
two solid lines in the $(\beta, \theta)$-plane (Fig.~4). 
The dot-dash lines represent the possible combination
of $\beta$ and $\theta$ resulting from the apparent separation
velocity of the two components, which is expressed by the relation
$\beta_\mathrm{sep}= (2 \beta \sin \theta)/(1-\beta^2 \cos^2
\theta)$. Finally, since we measure motion on both sides and we know
the source distance, we can directly solve for $\theta$ and $\beta$
as discussed by \cite{mir94}; this corresponds to the dashed ellipse
centered on $\theta=2.7^{\circ}$, $\beta=0.79$.

   \begin{figure}
   \centering
   \vspace{180pt}
   \includegraphics{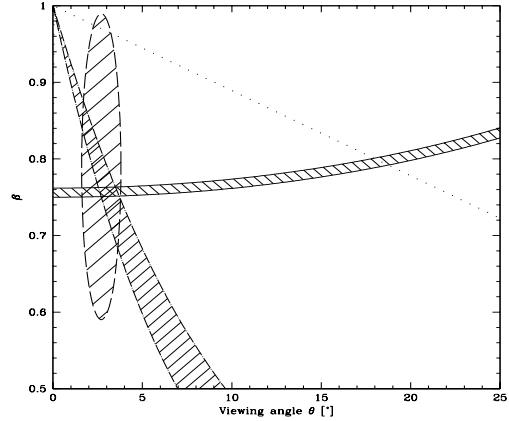}
      \caption{($\theta, \beta$) plane for NGC 4278, as discussed in the
text.}
   \end{figure}
 
In principle, one could also consider the brightness ratio between the
two components; however, as an effect of relativistic time dilation,
we are watching the components at different stages of evolution. Since
we know little on the time evolution of jet components, this hinders
the possibilty to apply the brightness ratio argument; in any case, 
$S_{N2}/S_{S2} > 1$, consistent with our
interpretation.
 
Based on the above analysis (Fig. 4), we find mildly
relativistic velocities of $\beta \sim 0.76$ ($\Gamma \sim 1.5$), and
an orientation close to the line-of-sight ($2^{\circ} \la \theta \la
4^{\circ}$). The resultant Doppler factor is $\delta \sim 2.7$; the
small viewing angle explains also the bendings visible in both jets,
as the amplification caused by projection effects of intrinsically
small deviations, which are common in low power radio
sources. 
 
\subsection{History of emission}
 
From the result of the modelfit, small to moderate apparent velocities
($\la 0.2\,c$) are found for the four jet components. Under the
assumption of constant velocity, we derive that they must have been
ejected from the core between 8.1 and 64.5 years before epoch 2000.652
(see Column [3] in Table 1).
 
However, these jet components are not to be confused with the hot
spots demarcating the end of the jet as found in more powerful CSOs by
\cite{ows98,pec00,gir03}. Therefore, a kinematic estimate of the real
age of the source is difficult, and that of $S1$ can only be taken as
a lower limit. The low brightness and large size of $N2$, as well as
the non detection of $N1$, suggest that components are continually
ejected from the core but that they soon disrupt, without being able
to travel long distances and form kpc scale lobes.

We do not expect that this source may evolve into a kpc scale radio
galaxy but that it will only periodically inflate slowly. 
The relatively low velocity jets discussed in
\S~\ref{jets} can not bore through the local ISM and escape, as shown
by the lack of hot spots, which are instead found in higher power
CSOs. This behavior has to be ascribed to a low power central engine,
which can not create highly relativistic jets.
 
   \begin{figure}
   \centering
   \vspace{200pt}
   \includegraphics{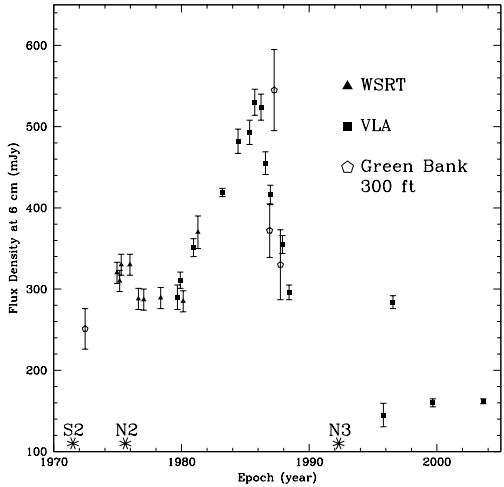}
      \caption{Light curve for NGC 4278 at 5 GHz}
   \end{figure}

In Fig. 5 we plot the flux density history for NGC~4278
at 6 cm, with data taken at the WSRT,
the Green Bank 300~ft radio telescope and the VLA. A previous plot was
published by \cite{wro91}, to which we add 12 points, from
observations obtained between 1972 and 2003. The light curve shows
that the source is variable, prone to both outbursts and low states. A
burst is certainly present around 1985, while in more recent years the
source has been showing less activity.
 
It is difficult to connect the burst with the ejection of new
components, both because of the uncertainties related to the age of a
single ``blob'', and the possible time lag between component ejection
and total flux enhancement. It is clear however that the source
presents a high degree of variability, possibly related to the
presence of an active nucleus.

For a more detailed discussion see Giroletti et al. (2004). 

\section{Conclusions}

In the present paper, we have presented new VLBA data for the
nearby ($d=14.9$ Mpc) LLAGN NGC~4278. Our data show a
two-sided emission on sub-parsec scales, in the form of twin jets
emerging from a central compact component ($T_B = 1.5 \times 10^9$ K),
in much a similar way to what happens in more powerful radio loud
AGNs. 
 
By comparison with previous data, we discover proper motion for
components in both jets, over a five years time baseline; we find low
apparent velocities ($\la 0.2\,c$) for the jet components and estimate
the epoch of their ejection as $10 - 100$ years before our
observations. Based on our analysis, we suggest that the north-west
side is the approaching side, and that the jets of NGC~4278 are mildly
relativistic with $\beta$ $\sim$ 0.75.
 
The central black hole in NGC~4278 is therefore active and able to
produce jets, which are responsible for the bulk of emission at radio
frequency in this LLAGN. However, the lifetime of
components of $< 100$ years at the present epoch and the lack of large
scale emission, suggest that the jets are disrupted before they reach
kpc scales.      %Future instruments, such as the New Mexico Array (EVLA phase\ 2), shall be able to add the sufficient short spacings necessary to see the c\omponents as they grow and fade.
 
The study of the flux density history at 6 cm between 1972 and 2003
shows a significant variability ($\ga 100\%$) on time scales of a few
years, which might be related to the ejection of new components. This
subject needs to be explored for other LLAGNs as well, as it can give
better insight about the state of the central black hole in these
sources.

\begin{acknowledgements}

MG thanks the NRAO for hospitality during his visit to Socorro when
much of this work was accomplished. The National Radio Astronomy
Observatory is operated by Associated Universities, Inc., under
cooperative agreement with the National Science Foundation. This
research has made use of NASA's Astrophysics Data System Bibliographic
Services and of the NASA/IPAC Extragalactic Database (NED) which is
operated by the Jet Propulsion Laboratory, Caltech, under contract
with NASA.  We thank also the
Italian Ministry for University and Research (MIUR) 
for partial support (grant 2003-02-7534).

\end{acknowledgements}

\end{document}